\documentclass[journal=jctcce,manuscript=article]{achemso}
\usepackage{siunitx}
\usepackage{dcolumn}\usepackage{bm}\usepackage{booktabs}
\usepackage[utf8]{inputenc}
\usepackage[T1]{fontenc}
\usepackage{etoolbox}

\DeclareSIUnit\angstrom{\textup{\AA}}
\DeclareSIUnit\atmosphere{atm}

\usepackage[version=3]{mhchem} 
\usepackage{hyperref}
\usepackage{multirow}
\usepackage{multicol}

\author{Rafael Bicudo Ribeiro}
    \affiliation[USP]{Institute of Physics, 
    University of São Paulo, 
    Rua do Matão 1731, 05508-090 São Paulo, São Paulo, Brazil}
\author{Henrique Musseli Cezar} \email{h.m.cezar@kjemi.uio.no}
\affiliation{ 
  Hylleraas Centre for Quantum Molecular Sciences
  and Department of Chemistry,
  University of Oslo,
  PO Box 1033 Blindern, 0315 Oslo, Norway }
\title{clusttraj: A Solvent-Informed Clustering Tool for Molecular Modeling}


\begin{document}

\begin{tocentry}

\includegraphics{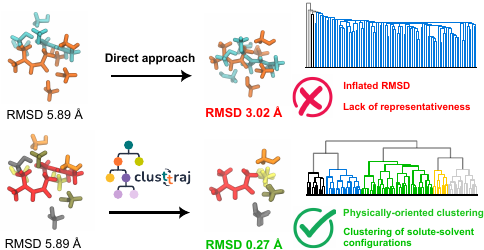}

\end{tocentry}

\begin{abstract}
Clustering techniques are consolidated as a powerful strategy for analyzing the extensive data generated from molecular modeling. 
In particular, some tools have been developed to cluster configurations from classical simulations with a standard focus on individual units, ranging from small molecules to complex proteins. 
Since the standard approach includes computing the Root Mean Square Deviation (RMSD) of atomic positions, accounting for the permutation between atoms is crucial for optimizing the clustering procedure in the presence of identical molecules.
To address this issue, we present the \texttt{clusttraj} program, a solvent-informed clustering package that fixes inflated RMSD values by finding the optimal pairing between configurations. 
The program combines reordering schemes with the Kabsch algorithm to minimize the RMSD of molecular configurations before running a hierarchical clustering protocol. 
By considering evaluation metrics, one can determine the ideal threshold in an automated fashion and compare the different linkage schemes available. 
The program capabilities are exemplified by considering solute-solvent systems ranging from pure water clusters to a solvated protein or a small solute in different solvents.
As a result, we investigate the dependence on different parameters, such as the system size and reordering method, and also the representativeness of the cluster medoids for the characterization of optical properties.
\texttt{clusttraj} is implemented as a Python library and can be employed to cluster generic ensembles of molecular configurations that go beyond solute-solvent systems. 
\end{abstract}

\section{\label{sec:intro} Introduction}

Classical simulations of molecular systems are widely spread in several research fields\cite{mitra2011}, enabling the modeling from gas-phase atoms towards complex systems such as ionic liquids\cite{maginn2009}, macromolecules solvated in biological environments\cite{karplus1980, haohao2022, pedebos2022} and various materials\cite{arash2021, chen2018, krishna2021}. 
With the development of high-performance computational packages, Molecular Dynamics (MD) and Monte Carlo (MC) methods became the main approaches at the atomistic and coarse-grained levels.\cite{allen-tildesley2018}
From the sampled configurations, thermodynamic properties can be computed, and the analysis of the trajectories may provide a valuable understanding of mechanisms dictated by free energy variations.
For instance, one can compare relative populations of conformers \cite{henrique2018}, track the denaturation of proteins by monitoring the radius of gyration \cite{lobanov2008} or even establish the preferential stacking between semiconductors in thin films.\cite{nfa-history2021}

Regardless of the application, the high volume of information contained in the trajectories benefits from data-driven analysis techniques.
Notably, Machine Learning (ML) methods have been employed to extract powerful insights by clustering similar configurations\cite{ttclust, quicksom, mdscan}. 
When grouping the configurations according to the distance between atoms, we can search for key features by comparing inter- and intra-cluster observations, but also select representative configurations.
These configurations can be used for several applications such as extracting the molecular configuration from the full wave function without evoking the Born-Oppenheimer approximation,\cite{Lang2024} or for identifying phase changes in transition metal nanoclusters.\cite{Cezar2017,Cezar2019}
Other important application is connected to reducing the overall cost when employing highly computationally demanding methods, \emph{e.g.}, in sequential quantum-mechanics/molecular-mechanics (s-QM/MM) calculations.\cite{kaline2008, Ribeiro2025}
By finding representative configurations, one can dramatically reduce the number of quantum chemistry calculations, making it feasible to use computationally demanding methods (\emph{e.g.}, multiconfigurational post-Hartree-Fock methods) that would be prohibitive if hundreds or thousands of configurations were considered.

Several methods have been proposed to cluster configurations from the Root Mean Square Deviation (RMSD) between snapshots of molecular dynamics trajectories.
A traditional approach involves binary implementations of quality threshold and Daura's algorithms to reduce the amount of RAM required to store the distance matrix\cite{campello2013, bitclust}. 
The former is implemented in programs such as VMD\cite{vmd1996} and GROMACS\cite{gromacs2015}, while the latter can be performed via BitClust\cite{bitclust} (or QTPy) and BitQT\cite{bitqt} packages. 
Furthermore, the graph-based Density Peaks formalism and the Self-Organizing Maps (SOM)-based algorithm, respectively implemented in the RDPeaks \cite{rdpeaks} and quicksom \cite{quicksom} codes, further reduce the computational demand, allowing the analysis of longer trajectories. 
Another widely employed strategy is the application of the hierarchical clustering scheme.\cite{Murtagh2017} 
Despite the high sensitivity to outliers, the hierarchical clustering algorithms stand out when the number of clusters is unknown\cite{shao2007}. 
A well-established implementation of the hierarchical clustering scheme is available in the TTClust\cite{ttclust} program, while further developments are implemented in the MDSCAN.\cite{mdscan}
The latter combines the Density-Based Spatial Clustering of Applications with Noise (DBSCAN) method with the hierarchical clustering scheme in an efficient implementation to remove the dependency on the pairwise similarity matrix and reduce overall memory consumption.\cite{campello2013}
Moreover, the MDANCE package\cite{chen_k-means_2024} provides a collection of tools with implementations of both $k$-means and hierarchical clustering approaches designed to cluster configurations of biophysical systems.\cite{chen_extended_2025, Chen2025.03.05.641742, chen_protein_2024, doi:10.1021/acs.jcim.5c00240}

\begin{figure}[htbp]
    \centering
    \includegraphics[width=240pt]{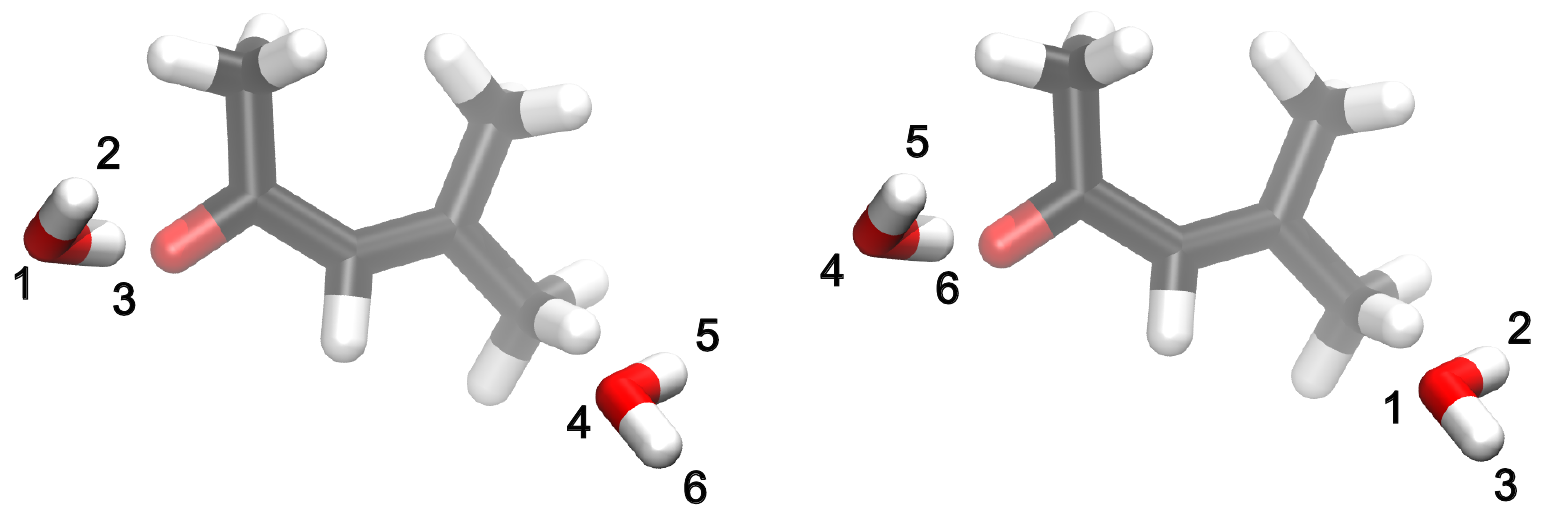}
    \caption{Illustration of two identical configurations with different atomic labels. The naive computation of the RMSD leads to a non-zero RMSD, despite the expected RMSD being zero in this case, since both water molecules are identical.}
    \label{fig:labels}
\end{figure}

Most of these packages are developed with a focus in the solute.
The use of clustering for trajectories that include the solvent remains relatively unexplored.
Considering solvent molecules is challenging because of the labeling of atoms when molecular configurations are represented in the computer memory.
This leads to possible problems due to the permutation between two identical molecules resulting in different RMSDs, despite the configurations being identical (see Figure~\ref{fig:labels}).
The high RMSD for configurations that are very similar in practice leads to artificial clusters when hierarchical clustering is performed.
Attempting to tackle this problem, Frömbgen \emph{et al.} \cite{frombgen2022} incorporated a hierarchical clustering algorithm into the TRAVIS\cite{travis2011} post-processing software to cluster liquid conformations.
However, the authors' goal was not necessarily to find representative configurations, and not many options regarding the hierarchical clustering were explored.

Here, we present \texttt{clusttraj}, an open source Python package that accounts for permutations to minimize the RMSD within the hierarchical clustering scheme. 
The package is available in the Python Package Index (PyPI), and can be installed with a simple \texttt{pip install clusttraj}.
The program is built on top of the RMSD package by Kromann\cite{Charnley2025}, uses efficient Python libraries such as NumPy\cite{harris2020array}, scikit-learn \cite{sklearn} and SciPy,\cite{2020SciPy-NMeth} and explores the embarrassingly parallel nature of the distance matrix calculation. 
The package is built with a focus on flexibility regarding the linkage methods within the hierarchical clustering, reordering of labels, the contribution of the solute and solvent, and which atoms are used for the RMSD computation.
To illustrate some of these possibilities, we analyze the clustering of water clusters of different sizes, a small solute (mesityl oxide) with different solvents, and a larger solute (lysozyme) in water.
We investigate metrics for the automatic detection of clusters, explore the different algorithms for hierarchical clustering and label reordering, and provide benchmarks of the performance.

\section{Implementation}
\label{sec:impl}
\subsection{RMSD and RMSD Matrix}
The implementation uses RMSD as a metric of similarity between two configurations.
The RMSD between the configurations $\mathbf{A} = \{ \mathbf{a}_1, \mathbf{a}_2, \dots, \mathbf{a}_N \}$ and $\mathbf{B} = \{ \mathbf{b}_1, \mathbf{b}_2, \dots, \mathbf{b}_N \}$, both with $N$ atoms, is defined as
\begin{equation}
    \label{eq:rmsd}
    \text{RMSD}(\mathbf{A}, \mathbf{B}) = \sqrt{\sum_{i=1}^N w_i \vert\vert \mathbf{a}_i - \mathbf{b}_i \vert\vert^2}
\end{equation}
where the $\mathbf{a}_i$ and $\mathbf{b}_i$ are the Cartesian coordinates for the atom of label $i$ (expressed in \si{\angstrom}), and $w_i$ is the weight of the $i$-labeled atom ($\sum_{i=1}^N w_i = 1$).
Usually, $w_i = 1/N$ for all $i$, but assigning different weights to each atom can be useful, as we show later. 
As illustrated in Figure~\ref{fig:labels}, the minimum value of RMSD is obtained when $\mathbf{A}$ and $\mathbf{B}$ are spatially aligned and the labels in each configuration are such that atoms of the same species in similar spatial positions correspond to the same label.

We use the package \texttt{rmsd}, available in PyPI, to align and reorder the labels.\cite{Charnley2025}
The package implements the spatial alignment of the configurations using the Kabsch algorithm\cite{kabsch1,kabsch2}.
Five methods are available for the reordering of the labels: the Hungarian algorithm\cite{Crouse2016}, an ordered distance criterion, FCHL19 atomic descriptors\cite{Christensen2020} with a Hungarian cost-assignment function (also known as QML algorithm), a distance to the center of mass approach, and the brute force method.
In principle, only the brute force method, which systematically considers all possible permutations of atoms of the same species, is guaranteed to find the minimum RMSD between two configurations.
However, in many circumstances, especially when the configurations are properly aligned, the other algorithms can also minimize the RMSD at a significantly lower computational cost (see Section~\ref{sec:time}, and Section~S3 in the SI).

\begin{figure*}
    \includegraphics[width=504pt]{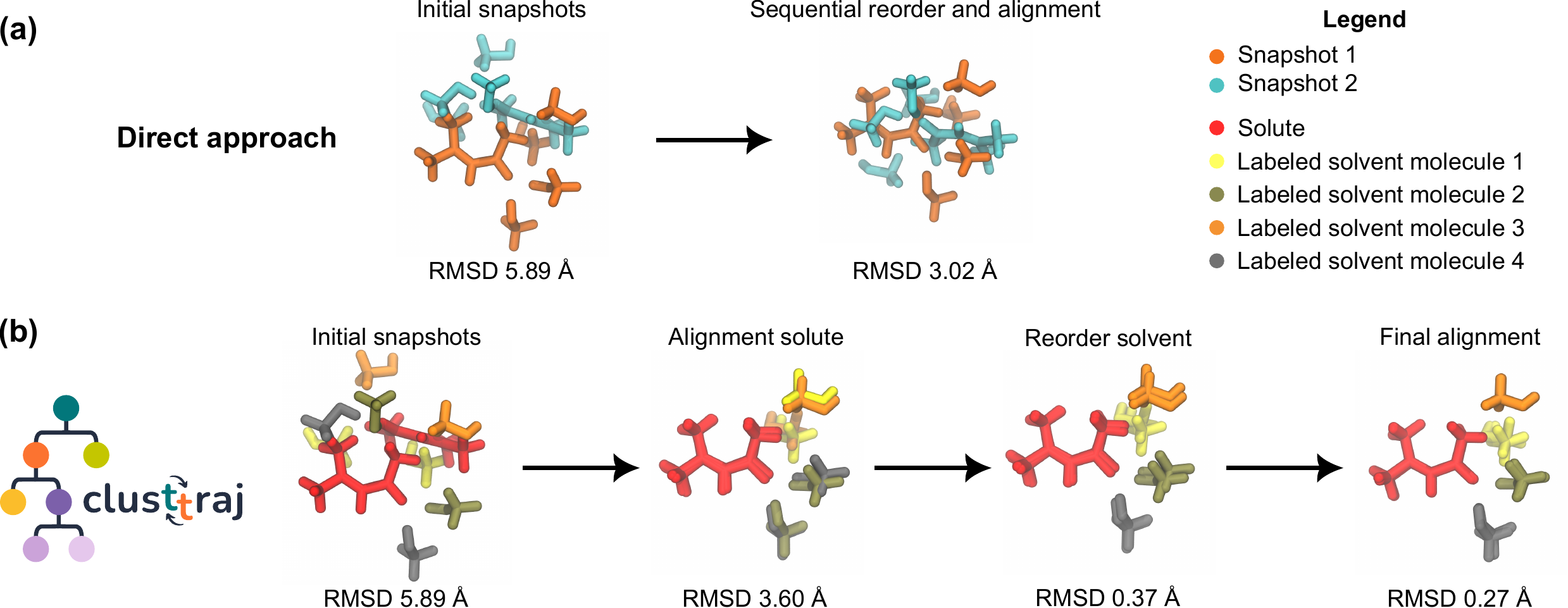}
    \caption{Steps illustrating the \texttt{clusttraj} strategy to find the minimum RMSD with the solute-solvent split. (a) RMSDs obtained with the simultaneous alignment and reorder of all the atoms, not distinguishing between solute and solvent atoms. (b) Strategy employed by \texttt{clusttraj}, with the RMSDs at each step.}
    \label{fig:clusttraj-steps}
\end{figure*}

The algorithm to find the minimum RMSD of the \texttt{rmsd} package first reorder the labels of $\mathbf{B}$, to then find the optimal rotation that aligns the two configurations and, finally, compute the RMSD using Equation~\ref{eq:rmsd}.
The reordering is performed first to ensure the Kabsch algorithm finds the correct rotation.
We call this the ``direct approach'', and illustrate it in Figure~\ref{fig:clusttraj-steps}a.
However, the reordering algorithms may depend on the initial distances between atoms of the same species to find the optimal labels, as in the case of the Hungarian algorithm.
Therefore, different strategies such as aligning the moments of inertia, or as we do in \texttt{clusttraj}, split the solute and solvent atoms, may improve the reordering of labels and lower the RMSD.

Based on the user options, \texttt{clusttraj} can consider a subset or all atoms in a configuration.
There are options to exclude all \ce{H} atoms from the clustering or manually select exclusions.
The exclusions are very flexible and can be passed to the command-line interface through ranges of atom labels in the initial trajectory.
By excluding atoms, one can reduce the number of possible permutations for the reordering, and also solve the problem of having to reassign labels for the hydrogen atoms in methyl groups.
These exclusions determine the size $N$ of the $\mathbf{A}$ and $\mathbf{B}$ matrices.

To compute the RMSD matrix, we first parse all $M$ configurations of the trajectory using OpenBabel.\cite{OBoyle2011}
By using OpenBabel, many standard chemical file formats are supported for both input and output of the aligned structures, such as \texttt{PDB}, \texttt{xyz}, \texttt{gro}, among others.
Snapshots are parsed one by one, and the entire trajectory is not loaded in memory at once.
The hierarchical clustering is performed based on the RMSD between pairs of the $M$ snapshots, which are combined to form the RMSD matrix.
The $M(M - 1)/2$ RMSDs between all the configurations in the trajectory are computed in parallel using the \texttt{multiprocessing} Python library.

\subsection{Solute-solvent split strategy}
The main difference between \texttt{clusttraj} and its alternatives is the solute-solvent strategy used to improve the clustering.
To exploit schemes that improve the search for the minimum RMSD with relabeling and the separation of dendrograms, the $N^\text{solute}$ atoms of the solute are treated separately from the atoms of the solvent.
This is relevant because often the solute molecule will have a greater importance than the solvent, so one can emphasize the best alignment of the solute while still accounting for solvent contributions.
\texttt{clusttraj} uses multiple strategies to account for these contributions.

Using different weights for the solute atoms, one can give a higher or lower importance to these contributions.
By default, the $w_i$ in Equation~\ref{eq:rmsd} are equal for all atoms in the configuration.
In \texttt{clusttraj}, it is possible to define $W^\text{solute} \in [0, 1]$, the total solute weight, so
\begin{equation}
    \label{eq:weights}
    w_i = 
    \begin{cases}
    \frac{W^\text{solute}}{N^\text{solute}} & \qquad i \leq N^\text{solute} \\
    \frac{1 - W^\text{solute}}{N - N^\text{solute}} & \qquad i > N^\text{solute}
    \end{cases}
\end{equation}
where we assume that the first $N^\text{solute}$ indexes correspond to the solute atoms.

The alignment and reordering are split in different steps for the solute and solvent, as shown in Figure~\ref{fig:clusttraj-steps}b.
This strategy takes advantage of a smaller search space for the label reordering and favors a better clustering with respect to the solute.
Also, since the reordering algorithms may depend on the Euclidean distance between the atoms of the same species, we first center the coordinates at the solute and perform a Kabsch alignment of the solute atoms only.
This often leads to good alignment even before reordering, especially when \ce{H} atoms are not considered, and it anyway serves as an approximate rotation to improve the initial RMSD.
From this initial rotation, to handle cases where the order of the solute atoms has changed, the reassignment of the solute labels is performed.
Based on this relabeling, a new Kabsch rotation guarantees optimal alignment between the two structures.

After these operations have been performed for the solute, we focus on the solvent molecules.
Excluding the solute atoms from the assignment problem, we relabel the atoms of the solvent molecules.
At this point, the configurations should be optimally aligned between the solutes, as they are in Figure~\ref{fig:labels}, and the relabeling of the solvent molecules is performed on a smaller subset of atoms and Cartesian coordinates that are similar for molecules in the same spatial region.
Under these circumstances, the reassignment algorithms can perform their best and possibly find the minimum RMSD.
Therefore, even algorithms that rely on some heuristics for relabeling, the case of the Hungarian or distance algorithms implemented in the \texttt{rmsd} package and used in \texttt{clusttraj}, usually work better with the reduced number of atoms in each step and after alignment.
A final Kabsch rotation including the solvent atoms can then optionally be performed before the final RMSD is returned.

A comparison between the direct approach, as implemented in the \texttt{rmsd}\cite{Charnley2025} package, where the solute and the solvent are treated simultaneously without distinction, and the \texttt{clusttraj} approach is shown in Figure~\ref{fig:clusttraj-steps}.
In this example, the RMSD found by the direct approach (\qty{3.02}{\angstrom}) is much higher than that found by \texttt{clusttraj}'s solute-solvent informed strategy (\qty{0.27}{\angstrom}) when using $w_i = 1 / N$.
Other strategies such as initial alignment of the principal inertia moments, implemented in the \texttt{rmsd} package, lower the RMSD in the direct approach (\qty{2.65}{\angstrom}), yet fail to obtain the minimum value found with the solute-solvent split strategy.

\subsection{Agglomerative Hierarchical Clustering}
In agglomerative hierarchical clustering a tree-like structure of nested clusters is built by iteratively merging data points based on similarity up until all clusters are merged into one.
Initially, each structure corresponds to its own cluster, and, based on a linkage algorithm, pairs of clusters are joined hierarchically \cite{Murtagh2017}.
The structure formed by this linkage is known as dendrogram, and gives a visual and quantitative representation of the distance between clusters.
Based on a cut at a certain tree height, the number of clusters and the members of each cluster are determined.

In \texttt{clusttraj}, the hierarchical clustering is performed using SciPy.\cite{2020SciPy-NMeth}
SciPy has many different algorithms implemented for linkage, such as single, average, complete, ward, among others.\cite{Mullner2011}
The clusters and separation between clusters are heavily determined by which algorithm is used.
\texttt{clusttraj} defaults to the ward linkage, which as we show in the results section, provides a good separation between clusters.

The threshold distance of the dendrogram is also another important quantity that determines the quality of the clustering.
If a too high threshold is used, the algorithm returns only very few clusters, while a too small threshold gives basically no clustering at all.
Some metrics can be used to optimize this threshold, such as the silhouette score (SS) \cite{rousseeuw1987}, the Calinski Harabasz score (CH) \cite{harabasz1974}, the Davies-Bouldin score (DB) \cite{bouldin1979} and the cophenetic correlation coefficient (CPCC) \cite{rohlf1962}.
In principle, one should aim to maximize the SS, which is a coefficient that ranges from \num{-1} to \num{1}, with positive values indicating that, on average, each point is closer to other points in the same cluster than to those in different clusters.
CH and DB are built around the cluster centroids, favoring spherical distributions, while CPCC consider heights from the hierarchical tree to assess the dissimilarities between the dendrogram and the original data.
These metrics are useful in helping determine the threshold, but a visual inspection of the similarity between cluster members is always necessary.
\texttt{clusttraj} can determine the threshold automatically based on the SS, or a manual threshold can be used.

\subsection{Output information}
After classification, \texttt{clusttraj} outputs information about the clustering process and the clusters.
The RMSD matrix is stored to save computational time when rerunning analysis with a different threshold or linkage algorithm, but also to allow analysis regarding the minimum RMSD distance between cluster members.
In addition to the RMSD matrix, the classification of each snapshot, the dendrogram, reduced-dimensionality plots, and classification evolution over time are saved.

For a visual analysis of the similarity of structures between clusters, \texttt{clusttraj} also outputs one trajectory file for each cluster, containing the best superposition of each cluster member with the medoid.
We define the medoid as the structure that has the smallest sum of RMSD to all other cluster members, and consider it the representative structure of the cluster for analysis in the results section. 

\section{Example Applications}
\label{sec:examples}
\subsection{Lysozyme protein}
\label{sec:lyso}
Since clustering methods are often applied in the context of bioinformatics, we investigated the lysozyme protein conformations in solution.
An NVT thermalization of the 1AKI crystal structure lysozyme protein solvated in water and neutralized with \ce{NaCl} at a concentration of \qty{0.15}{\mol\per\liter} was performed for \qty{10}{\ns}, employing the leap-frog integrator with a time step of \qty{2}{\fs}. 
Hydrogen bonds were constrained and the temperature was set at \qty{300}{\kelvin} using the modified Berendsen thermostat \cite{parrinello2007} with a dumping constant of \qty{0.1}{\ps}. 
The OPLS-AA\cite{jorgensen1988, jorgensen1996} and TIP3P\cite{jorgensen1983} force fields were employed for the protein and water, respectively.
From the trajectory file, \num{100} snapshots were extracted to apply the clustering procedure.<
The simulation was performed with GROMACS\cite{gromacs2015} and the GROMACS toolkit was combined with the functionalities from MDAnalysis\cite{mdanalysis} package to modify trajectory files and analyze the clustering results.

\begin{figure}[htb]
    \centering
    \includegraphics[width=240pt]{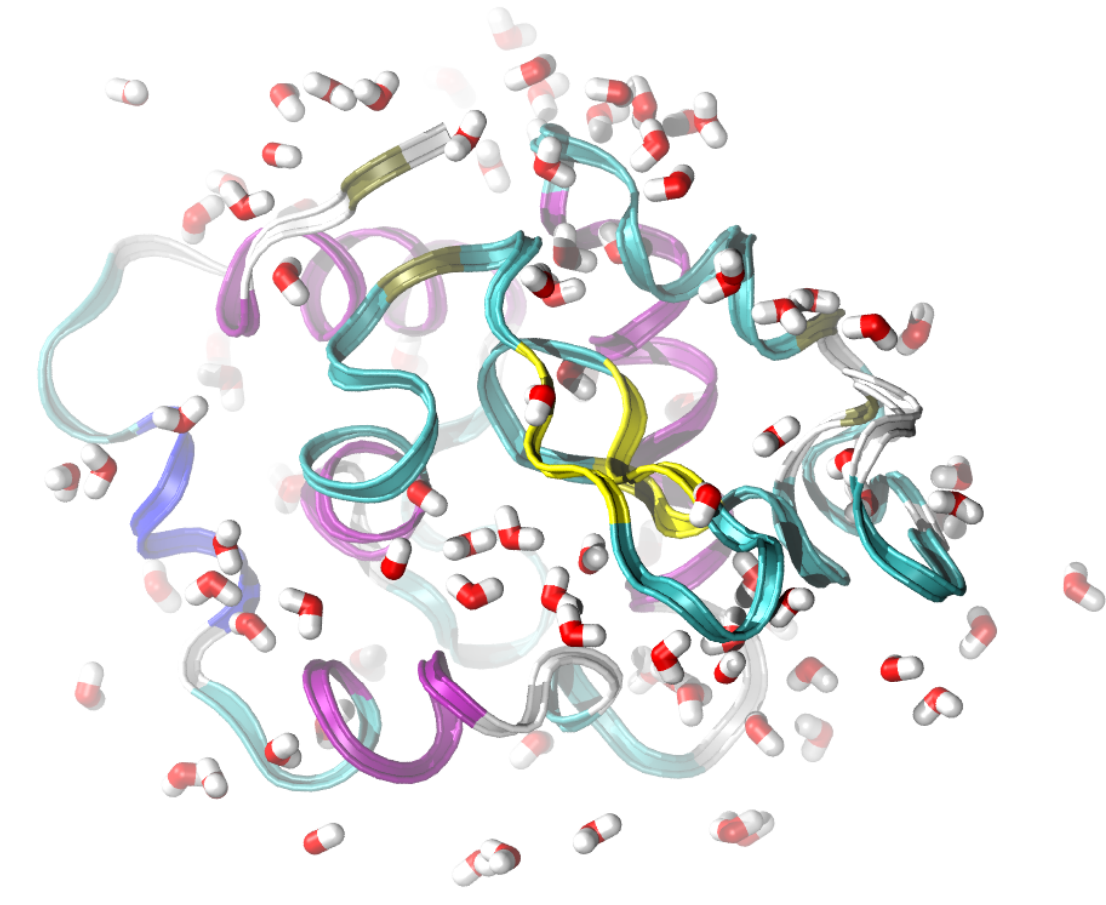}
    \caption{Snapshot of the first frame of the lysozyme trajectory. The protein is solvated by \num{100} water molecules.}
    \label{fig:lyso}
\end{figure}

\begin{figure}
    \centering
    \includegraphics[width=240pt]{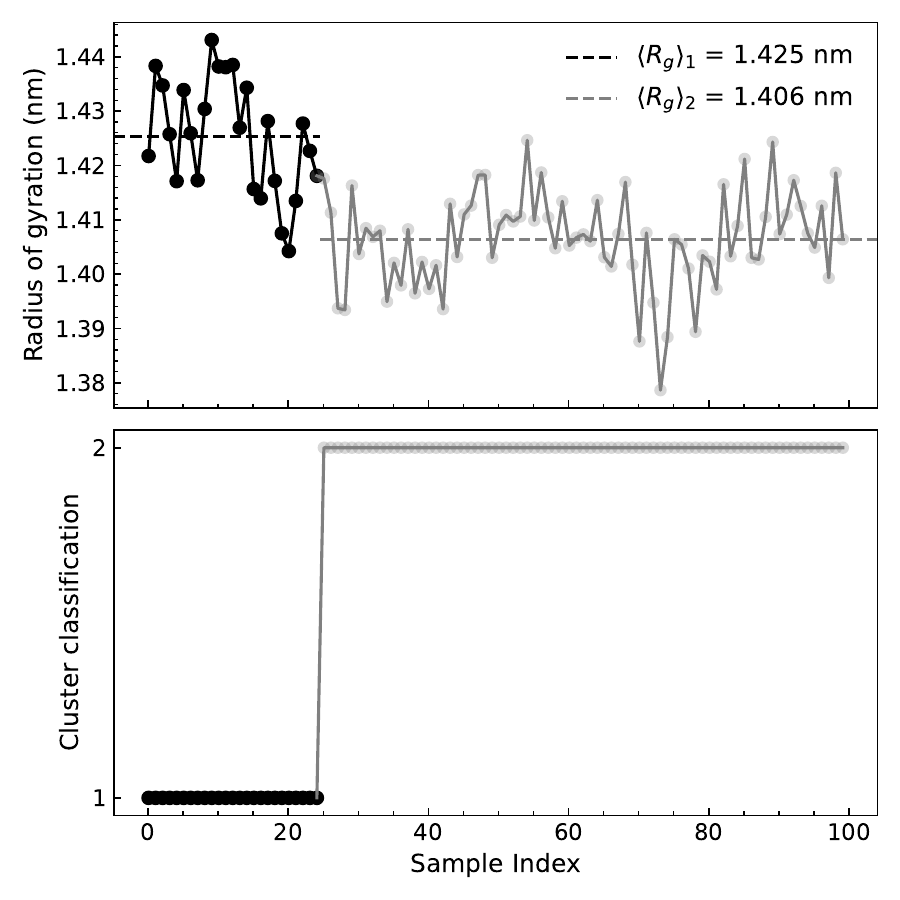}
    \caption{Radius of gyration (top) compared with the cluster evolution (bottom) for the 100 configurations of the lysozyme protein solvated by 100 water molecules.}
    \label{fig:lyso_gyr}
\end{figure}

At first, we investigated the solvent distribution around the amino acids by including the 100 water molecules whose atoms were closest to any protein atom, as illustrated by Figure~\ref{fig:lyso}.
We combine the Hungarian algorithm for label reordering with the Ward variance minimization method to calculate the distance between clusters. 
The RMSD threshold was chosen to maximize the SS, and the hydrogen atoms were not included, resulting in two clusters with \num{25} and \num{75} samples each.
Since these snapshots were extracted from a trajectory file, we can track the time evolution of the solvated protein conformation. 
The first \num{25} configurations belong to cluster 1, and the remaining belong to Cluster 2.
Considering that the RMSD of atomic positions is provided as input for the clustering procedure, one should expect changes in the structural properties of the snapshots from different clusters.
In particular, as shown in Figure~\ref{fig:lyso_gyr}, the cluster shift is related to the radius of gyration of the protein.
From the top panel, we can observe higher values for Cluster 1 and a difference of \qty{0.019}{\nano\meter} between each cluster's average radius of gyration, corresponding to a substantial \qty{29.7}{\percent} of the observed range of \qty{0.064}{\nano\meter}.
Therefore, the clustering procedure allowed the configurations to be grouped according to the protein packing while still considering the interaction with the solvent.

One can also focus on different properties by changing the atoms in the configuration file. 
For example, by considering only the lysine amino acid with the closest water molecules, we could identify differences in the number of hydrogen bonds of configurations from different clusters. 
See the Supporting Information (SI) for more details.

\subsection{Water}
\label{sec:water}
For the second example, we performed a molecular dynamics simulation of \num{500} water molecules in the NVT ensemble, with a density of \qty{1}{\g\per\cm^3}.
The equations of motion were integrated using the leap-frog algorithm with a timestep of \qty{0.1}{\fs}.
The temperature was kept constant at \qty{300}{\kelvin} using the modified Berendsen thermostat\cite{parrinello2007} and a damping constant of \qty{0.1}{\ps}.
We used the TIP3P force field to describe the interactions\cite{jorgensen1983} and the simulation was performed using GROMACS\cite{gromacs2015} software.
From the \qty{1}{\ns} simulation, we extracted \num{100} snapshots and a random water molecule (the solute) was chosen as a reference to generate the input file for the clustering procedure.
We varied the number of nearest neighbor molecules (solvent molecules) from 1 to 100 to investigate the system size dependence.
The Hungarian algorithm and Ward variance minimization method were employed to reorder the atoms and to compute the distance between clusters within the linkage scheme, respectively.
The threshold RMSD was chosen to maximize the silhouette score, yielding the results presented in Table \ref{tab:waters}.

\begin{table}[htb]
    \caption{Clustering results for different number of neighbor water molecules.}
    \begin{tabular}{ccccp{1.75cm}}
        \toprule
        \# of waters & \# of clusters & RMSD (\AA) & Silhouette score & RMSD matrix sum (\AA) \\         \midrule
        1 + 1 & 3 & 1.822 & 0.413 & 1720 \\         1 + 2 & 8 & 1.670 & 0.261 & 3177 \\         1 + 3 & 4 & 4.031 & 0.261 & 4547 \\         1 + 4 & 41 & 1.290 & 0.153 & 4404 \\         1 + 5 & 32 & 1.818 & 0.148 & 4843 \\         1 + 6 & 45 & 1.704 & 0.132 & 5057 \\         1 + 7 & 47 & 1.814 & 0.114 & 5133 \\         1 + 8 & 39 & 1.934 & 0.106 & 5102 \\         1 + 9 & 45 & 1.873 & 0.095 & 5103 \\         1 + 10 & 38 & 1.928 & 0.073 & 5080 \\         1 + 20 & 50 & 1.877 & 0.040 & 4987 \\         1 + 30 & 46 & 1.905 & 0.028 & 4948 \\         1 + 40 & 41 & 1.949 & 0.024 & 4921 \\         1 + 50 & 38 & 1.944 & 0.027 & 4927 \\         1 + 100 & 2 & 2.536 & 0.015 & 4758 \\         \bottomrule
    \end{tabular}
    \label{tab:waters}
\end{table}

Although the RMSD threshold can be chosen by hand, one should consider the evaluation metrics to improve the clustering procedure. 
Despite having positive SS values for all the systems presented in Table \ref{tab:waters}, there is an explicit dependency on the system size.
The clustering quality decreases as we add more atoms, reflecting the broader variety of configurations due to increasing the system's degrees of freedom.
This conformational diversity is connected to the RMSD between snapshots.
Since low RMSD indicates a higher similarity between configurations, the sum of RMSD between all snapshots can be considered to assess the quality of the reordering and minimization procedures.
As shown in the last column of Table \ref{tab:waters}, the sum of RMSDs grows as we increase the number of solvent waters but eventually converges. 

Despite the high correlation between SS and the sum of RMSDs (Pearson's correlation coefficient of -0.847 for values in Table \ref{tab:waters}), the assessment of cluster quality should not be the only metric.
For instance, when considering 1 + 7 and 1 + 100 systems, the SS indicates a better clustering for the smaller system while the RMSD sum favors the larger one.
This contrast can also be observed for the clustering of 1 + 2 and 1 + 3 systems, which share the same SS but have significantly different RMSD sums of \qty{3177}{\angstrom} and \qty{4547}{\angstrom}, respectively.
Therefore, the clustering quality does not necessarily reflect the quality of the reordering and minimization procedures.

Examining the RMSD thresholds can also provide valuable information regarding the inter- and intra-cluster similarity between configurations. 
For 1 + 2 and 1 + 3 systems, SS was maximized using  \qty{1.670}{\angstrom} and \qty{4.031}{\angstrom} thresholds, respectively. 
This distinction indicates that configurations belonging to the same cluster are more diverse for the larger system, suggesting that the medoid is less representative of the overall cluster.
On the other hand, having more clusters may fail to reduce the complexity, requiring a balanced description between the number of clusters and the evaluation metrics.
This highlights the importance of fine-tuning the RMSD threshold to perform physically accurate clustering.

\subsection{Mesityl oxide}
\label{sec:mox}
For the final example, we studied the mesityl oxide (MOx) [\ce{(CH3)2C=CHC(=O)CH3}, 4-methyl-3-penten-2-one] molecule solvated in methanol and in acetonitrile. 
The solute and the four closest solvent molecules were explicitly considered, and the configurations were extracted from Configurational Bias Monte Carlo (CBMC) simulations performed with DICE.\cite{Cezar2020a}
Simulations were performed in the NPT ensemble with $P=\qty{1}{\atmosphere}$ and $T=\qty{300}{\kelvin}$, with one MOx molecule and \num{800} solvent molecules.
The internal degrees of freedom of the solute were sampled with the CBMC algorithm, while the solvent molecules are kept rigid.
More details of the simulation conditions are given in a previous work.\cite{Cezar2020b}

\subsubsection{Methanol}
\label{subsec:mox_methanol}
\begin{figure}[htbp]
    \centering
    \includegraphics[width=240pt]{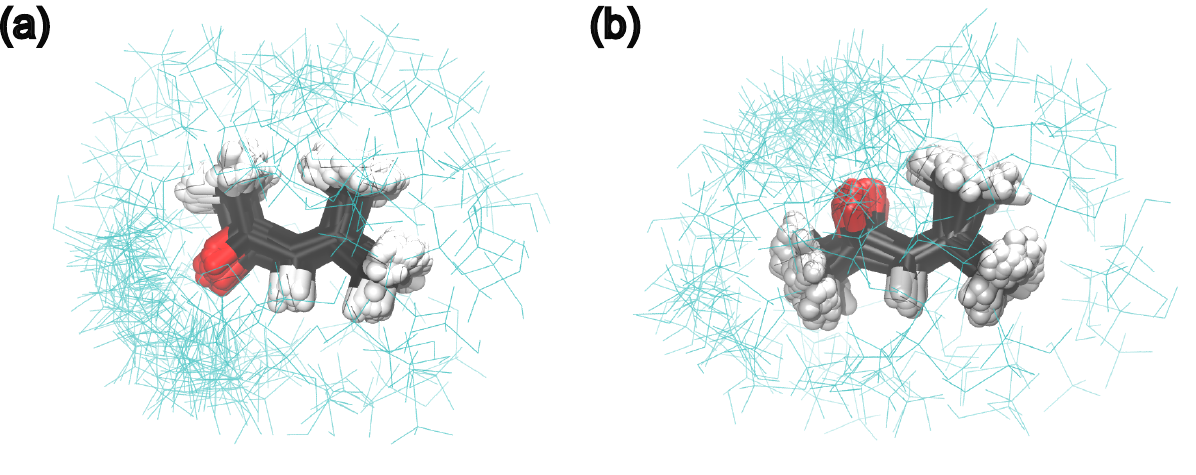}
    \caption{Superposition of MOx configurations separated in two clusters with (a) \emph{anti} and (b) \emph{syn} conformers. Methanol molecules are shown as cyan lines for better visualization.}
    \label{fig:mox}
\end{figure}

Considering a threshold of \qty{10}{\angstrom}, we obtained \num{3} clusters with significant structural differences between the configurations, as shown in SI.
However, more interesting results were obtained when changing the atom weights for solute and solvent molecules. 
Specifically, by excluding the \ce{H} atoms and increasing the weight of solute atoms to \qty{90}{\percent} of the total RMSD, we obtained two clusters that separate \emph{syn} and \emph{anti} conformations, as shown in Figure~\ref{fig:mox}. \\

Shao et al.\cite{shao2007} investigated how different algorithms behave for clustering solute configurations obtained from molecular dynamics trajectories and found that the average linkage algorithm (also known as UPGMA) performed better than its alternatives.
To investigate the clustering sensitivity towards the linkage method with the solute-solvent splitting of \texttt{clusttraj}, we compared the different approaches available to generate a fixed number of \num{5} clusters and the results are shown in Table \ref{tab:metrics}.
We evaluated all clustering metrics implemented in \texttt{clusttraj}, namely, the SS, CH, DB and CPCC.
In practice, our goal is to maximize SS, CH and CPCC while minimizing DB, which is not trivial according to Table~\ref{tab:metrics}.
Since the CH score considers the ratio between cluster separation and dispersion of configurations within the same cluster, the maximization of this coefficient agrees with the visual analysis of the dendrogram, disfavoring the use of simple, centroid and median distances (see SI).
In this sense, our results are consistent with the literature\cite{shao2007}, as average and ward distance methods are commonly used for hierarchical clustering of single molecule configurations from MD trajectories.

\begin{table}[htb]
    \caption{Evaluation metrics of hierarchical clustering with different distance methods. Root-Mean Square Deviation (RMSD), silhouette score (SS), Calinski Harabasz score (CH), Davies-Bouldin score (DB) and Cophenetic correlation coefficient (CPCC).}
    \begin{tabular}{cccccc}
        \toprule
        Method & RMSD (\AA) & SS & CH & DB & CPCC \\
        \midrule
        Single & 1.98 & -0.068 & 1.191 & 1.223 & 0.281 \\
        Complete & 4.40 & 0.056 & 8.555 & 2.828 & 0.371 \\
        Average & 3.10 & 0.081 & 6.285 & 1.739 & 0.544 \\
        Weighted & 3.10 & 0.084 & 7.107 & 2.220 & 0.469 \\
        Centroid & 2.43 & 0.060 & 3.186 & 1.291 & 0.431 \\
        Median & 2.40 & -0.002 & 3.043 & 1.305 & 0.452 \\
        Ward & 6.50 & 0.107 & 10.110 & 2.087 & 0.419 \\
        \bottomrule
    \end{tabular}
    \label{tab:metrics}
\end{table}

The usefulness of clustering configurations is also extendable to investigating quantum properties. 
In this case, we are interested in the solvatochromic shift of the absorption spectrum. 
The calculations reported in a previous work\cite{henrique2018} show a shift in the convoluted absorption spectrum when methanol molecules are included in the quantum region.
Within the s-QM/MM method, an ensemble of configurations is extracted from uncorrelated snapshots of classical simulations to perform the quantum calculations. 
The number of configurations required to achieve convergence varies according to the property. 
Typically, around \num{100} snapshots are required to properly access the configurational space, resulting in a computational demand depending on the level of calculations.
In this context, the benefit of clustering becomes evident when we improve the inter-cluster configuration variety by reducing the RMSD threshold to obtain \num{4} clusters of MOx and methanol molecules.
Considering the medoid configurations from each cluster, we obtain an average excitation energy of \qty{243.75}{\nm}.
Given the estimated error of $\sigma=\qty{100}{\cm^{-1}}$ (\qty{0.591}{\nm} for this wavelength), this result is compatible within a \qty{1.3}{\sigma} interval with the actual value of \qty{242.98}{\nm} determined from the \num{100} configurations.
However, if we randomly select, for instance, \num{10} sets with \num{4} configurations in each, we obtain an average excitation energy of \qty{241.44}{\nm} for the brightest excited state, deviating by \qty{2.6}{\sigma} from the convoluted value.
Considering the normal distribution, a \qty{1.6}{\sigma} deviation has a probability of approximately \qty{19.7}{\percent}, meaning it is relatively common and could easily occur due to random variation. In contrast, a \qty{2.6}{\sigma} deviation has a probability of only \qty{0.95}{\percent}, making it much rarer and more statistically significant.
Therefore, in contrast to the random sampling of configurations, the medoid approach provides a consistent solvatochromic shift of the brightest absorption peak and drastically reduces the computational cost. 
By only considering \num{4} configurations instead of \num{100}, we can also afford a higher computational cost per calculation, producing more reliable results, e.g., by expanding the quantum region or refining the level of theory employed.

\subsubsection{Acetonitrile}
\label{subsec:mox_aceto}
Similarly to the study in methanol, we performed the clustering of 100 configurations comprising a single MOx and \num{4} acetonitrile molecules. 
The clustering scheme included all hydrogen atoms, and we fixed an RMSD threshold of \qty{4.0}{\angstrom}. 
The Hungarian reorder algorithm was employed with $W^\text{solute}=0.9$, resulting in \num{4} clusters with populations \num{22}, \num{39}, \num{22} and \num{17}, respectively.

\begin{figure}[htbp]
    \centering
    \includegraphics[width=240pt]{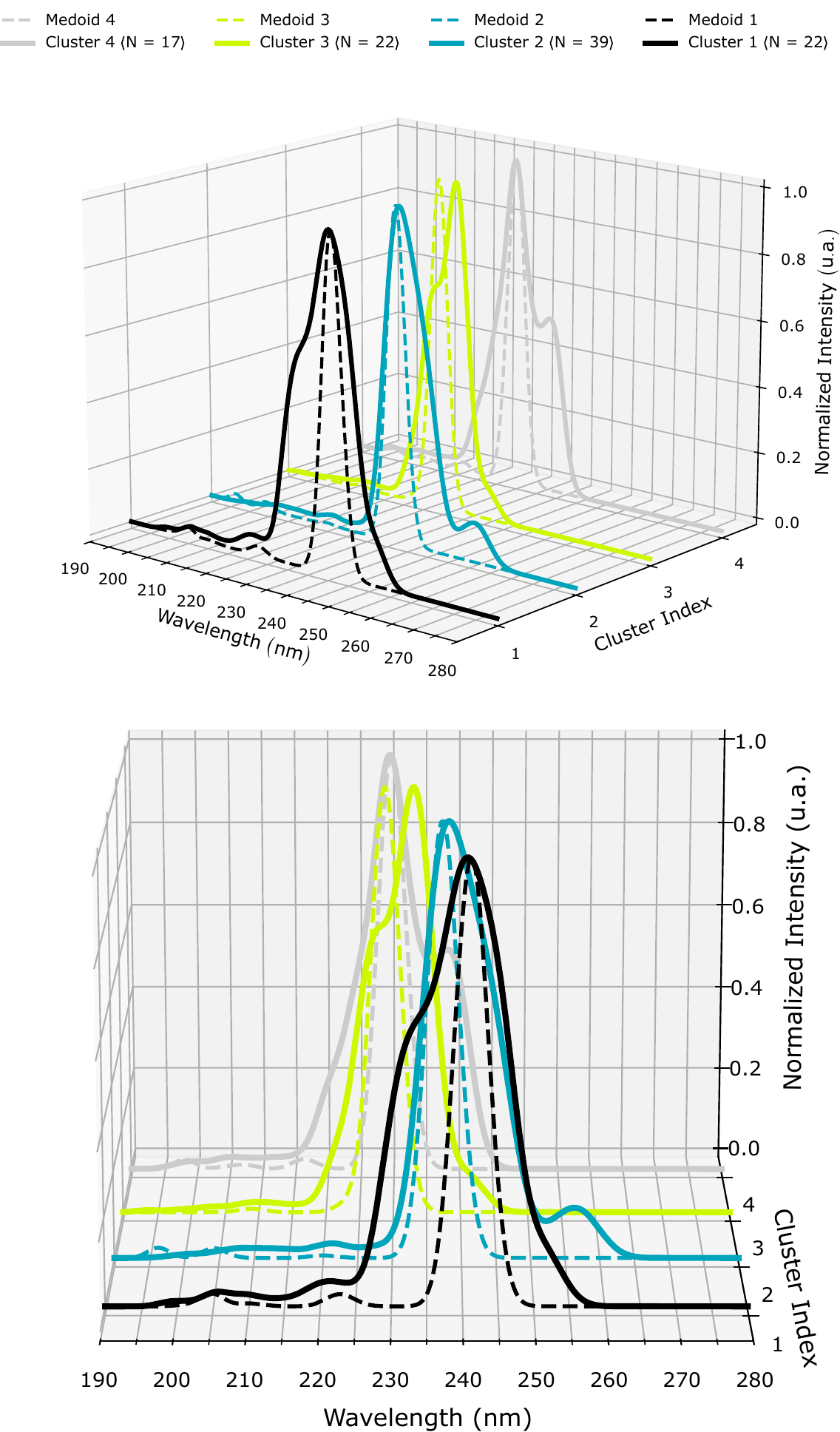}
    \caption{UV-vis absorption spectra of MOx and the 4 closest acetonitrile molecules. Dotted lines are the medoids' spectra only, and solid lines correspond to the averaged spectrum over all 22, 39, 22 and 17 configurations from clusters 1, 2, 3 and 4, respectively.}
    \label{fig:spec_mox_aceto}
\end{figure}

As an extension of the previous example, we compared the UV-vis absorption spectra of the medoid configurations against their respective average cluster spectrum. 
The spectrum was determined by fitting Gaussian functions to the first \num{6} singlet excited states, resulting in the curves shown in Figure~\ref{fig:spec_mox_aceto}.

Despite the enlargement in the peak width that arises from the different configurations, the medoid spectra (dotted lines) generally matched the brightest excitation energy.
For clusters \num{1}, \num{2} and \num{4}, the medoid peak matches the brightest peak, while for cluster \num{3}, it corresponds to the second brightest peak. 
Since the differences in excitation energies and oscillator strengths come from the geometry changes between configurations, the medoid bypasses detailed shapes, such as the shoulder-like structure observed around \qty{230}{\nm} for the averaged spectrum of Cluster \num{1}.
Additional information concerning the solvent contribution to the desired excitation can be incorporated into the scheme by changing the weight of solute atoms.

This example is also important to illustrate the role of the RMSD threshold in the clustering procedure.
Despite the use of metrics such as the SS, CH, DB, and CPCC to evaluate clustering quality and even provide automatic thresholds for the RMSD, user assessment of the clustering remains extremely important.
While decreasing the RMSD threshold increases the number of clusters, it tends to reduce the variance between configurations in the same clusters, improving the representativeness of the medoid configuration.
The spectra in Figure~\ref{fig:spec_mox_aceto} show that each medoid contributes with different excitations, exhibiting maximum absorption at distinct wavelengths.
Consequently, the quality of the overall spectra obtained from the combination of the individual representatives depends on the RMSD threshold.
Few clusters will tend to have poor representative structures, while having many clusters will fail the purpose of performing the clustering.
The appropriate balance can only be determined by the user, as it is both system-dependent and application-dependent.

\subsection{Computational cost}
\label{sec:time}
As described in Section~\ref{sec:impl}, the program is developed on top of well-optimized Python libraries such as numpy and scikit-learn. 
However, the additional steps required to reorder and permute identical molecules can be very costly, making the performance comparison with other clustering packages unfeasible.

Regardless of the options used for the analysis, one must compute the RMSD matrix for all clustering schemes considering the RMSD of atomic positions.
For a trajectory with $M$ configurations, there are $M(M-1)/2$ pairwise RMSDs, resulting in a computational complexity scaling of $\mathcal{O}(M^2)$.
To find the optimal pairing between two configurations, we employ the Kabsch algorithm \cite{kabsch1, kabsch2} that generally scales linearly with the number of atoms, $N$.\cite{kabsch3}
However, the most computationally demanding part relates to the reordering step. 
In a simplified picture, considering all possible atomic permutations between atoms has a drastic scaling of $\mathcal{O}(N!)$. 
Although a natural strategy, this brute-force approach is unfeasible for larger systems, requiring other techniques such as the distance or Hungarian algorithms.
The former reorders the atoms concerning the distance to the system centroid, yielding an approximate solution with a computational complexity of $\mathcal{O}(N\log{N})$.\cite{sedgewick1978}
On the other hand, the latter describes the reordering process in terms of a linear assignment problem and finds the exact solution that typically scales with $\mathcal{O}(N^3)$.\cite{tomizawa1971, karp1972}

To investigate the \texttt{clusttraj} performance, we compared the computational demand for each reordering method with different system sizes.
We considered the same trajectory files of the first example, comprising a water molecule solvated by 1, 3, 10 or 100 water molecules.
All calculations used \num{4} threads in an Apple M2 processor.
Results are shown in Table~\ref{tab:time_consumption}.

\begin{table}[htbp]
    \caption{Time consumption of different reordering algorithms for clustering 1 water solvated by 1, 3, 10, 20 or 100 water molecules. The results were averaged over 10 independent simulations.}
    \begin{tabular}{cccc}
        \toprule
        Reordering algorithm & System & RMSD matrix (s) & Total time (s) \\
        \midrule
        & 1 + 1 & 2.17(0.28) & 4.14(0.42) \\
        Hungarian & 1 + 10 & 2.23(0.19) & 3.95(0.19) \\
        & 1 + 100 & 16.17(0.67) & 18.29(0.72) \\
        \hline
        \multirow{2}{*}{Brute force} & 1 + 1 & 2.34(0.17) & 4.15(0.16) \\
        & 1 + 3 & 38.14(0.87) & 40.13(0.87) \\
                \hline
        & 1 + 1 & 2.05(0.20) & 3.79(0.25) \\
        Distance & 1 + 10 & 2.27(0.15) & 3.99(0.19) \\
        & 1 + 100 & 7.30(1.20) & 9.71(1.53) \\
        \hline
        & 1 + 1 & 2.50(0.16) & 4.29(0.21) \\
        QML & 1 + 10 & 16.34(0.27) & 18.17(0.32) \\
        & 1 + 20 & 118.11(8.04) & 119.97(8.17) \\
        \bottomrule
    \end{tabular}
    \label{tab:time_consumption}
\end{table}

In all cases, the most expensive part concerns the RMSD matrix calculation, which includes the reordering scheme and the Kabsch algorithm. 
As expected, for practical purposes, the factorial scaling of the brute force algorithm limits the procedure to systems with up to dozens of atoms shared among identical molecules. 
On the other hand, the distance approach is the fastest one, allowing the treatment of larger systems and including more configurations. 
However, this approximated method yields worse metrics than the others, changing the number of observations per cluster and even the optimal number of clusters. 
For example, when considering the \num{1} + \num{100} system, the distance method produces two clusters with \num{42} and \num{58} observations, maximizing the SS to \num{0.052} with a RMSD matrix sum of \qty{50875}{\angstrom}. 
Considering the same set of configurations, the Hungarian algorithm produces \num{10} clusters with an SS of \num{0.205} and a RMSD matrix sum of \qty{31132}{\angstrom}. 
These differences between the distance and Hungarian algorithms are observed for all the other evaluation metrics and for the three systems considered in Table \ref{tab:time_consumption}.
See sections S3 and S4 of the SI for a detailed comparison between the quality of each reordering algorithm and a illustrative comparison with a standard clustering package, respectively.

\section{Conclusions}
\label{sec:concl}
We presented a novel tool for clustering molecular configurations within the hierarchical clustering approach. 
Given the importance of considering the permutation of identical molecules to improve cluster formation, we developed the \texttt{clusttraj} program, an open-source tool that allows the clustering of a new range of systems within a simple command line interface.

Following the standard approach, molecular configurations are clustered by considering the RMSD between atomic positions. 
However, our procedure includes an initial reordering scheme to find the optimal pairing, avoiding the spurious increase of the RMSD due to the permutation of identical molecules, enabling, for example, the clustering of solute-solvent systems. 
Different reordering algorithms are available, from the refined treatment of all possible permutations to approximate methods that enable the study of larger systems with extensive ensembles of configurations. 
In particular, the Hungarian algorithm presented the best cost-benefit among the methods available and was employed throughout the examples. 

By considering representative systems, we consistently varied the proportion between identical and nonidentical molecules to investigate the clustering performance.
We obtained higher evaluation metrics when considering the average and Ward variance minimization linkage methods for computing the inter-cluster distance.
Although no direct relationship was observed between performance and system size, clustering of the large lysozyme system captured structural differences, yielding two clusters with different protein gyration radii. 
This behavior not only indicates a correlation between the clustering procedure and the structural properties but also reproduces the results expected for the analysis of isolated molecules by capturing the solute properties despite the presence of solvent.

For the smaller systems, we observed that optimizing clustering metrics does not necessarily result in minimizing RMSD between configurations. 
However, by adjusting the weight of the solute atoms, we successfully separated the \emph{syn} and \emph{anti} conformations of a MOx molecule in methanol while maximizing the SS. 
Furthermore, by increasing the RMSD threshold to produce \num{4} clusters, we obtained a good agreement between the solvatochromic shift of the UV-vis absorption spectrum convoluted over \num{100} configurations and determined only from the cluster medoids. 
A similar trend was observed when considering MOx solvated in acetonitrile. 
For each cluster, the medoid configuration was able to reproduce at least one of the two brightest peaks of the absorption spectrum convoluted over all configurations of the same cluster. 
Therefore, considering the medoid configuration is a solid method for selecting representative configurations that can improve the comparison while drastically reducing the computational cost.
Since \texttt{clusttraj} provides one solution for clustering identical molecules, we present a valuable tool that covers traditional solution-based clustering applications and can leverage the molecular modeling of complex systems. 

\section{Data Availability Statement}

\noindent The \texttt{clusttraj} code is available free of charge on \url{https://github.com/hmcezar/clusttraj} under the GPL-3.0 license. The trajectory files and the \texttt{clusttraj} output of all the examples presented in the manuscript can be found in the Norwegian Research Infrastructure Services (NIRD) research data archive on \url{https://dx.doi.org/10.11582/2025.2ab1vwjm} .

\section{Conflicts of interest}
    
The authors declare no conflict of interest.

\begin{acknowledgement}
R.B.R. acknowledges financial support from CAPES (Grant N$^\text{o}$ 88887.644651/2021-00) and FAPESP (Grant N$^\text{o}$ 2022/04379-3). H.M.C. thanks the support of the Research Council of Norway through the Centre of Excellence \textit{Hylleraas Centre for Quantum Molecular Sciences} (Grant N$^\text{o}$ 262695) and Sigma2 -- the National infrastructure for high-performance computing and data storage in Norway (grant numbers NN4654K and NS4654K).
\end{acknowledgement}

\begin{suppinfo}
Hydrogen bond analysis for clusters of lysine amino acid solvated in water. Clusters of mesityl oxide (MOx) solvated in methanol considering a different RMSD threshold. Dendrograms from the clustering of MOx solvated in methanol using different linkage methods.
\end{suppinfo}

\bibliography{ref}

\end{document}


\begin{center}
    \thispagestyle{empty}
    \Large{$\boldsymbol{\textsc{Supporting Information}}$} 
    
    \vspace{3em}
    
    \Large{\textbf{clusttraj: A Solvent-Informed Clustering Tool for Molecular Modeling}}
    
    \vspace{2.5em}
    
    \large\text{Rafael Bicudo Ribeiro$^1$, Henrique Musseli Cezar$^{2*}$}

    \vspace{2.5em}
    
    \textit{\small
        \noindent $^1$ Institute of Physics, University of São Paulo, Rua do Matão 1731, 05508-090 São Paulo, São Paulo, Brazil \\
        $^2$ Hylleraas Centre for Quantum Molecular Sciences and Department of Chemistry, University of Oslo, PO Box 1033 Blindern, 0315 Oslo, Norway
    }

    \vspace{1em}

    \small{
        $^*$\faEnvelope: h.m.cezar@kjemi.uio.no \\
    }
    
\end{center}

\clearpage

\tableofcontents

\clearpage

\section{Lysine solvated in water}

\begin{figure}[htbp]
    \centering
    \includegraphics[width=0.95\linewidth]{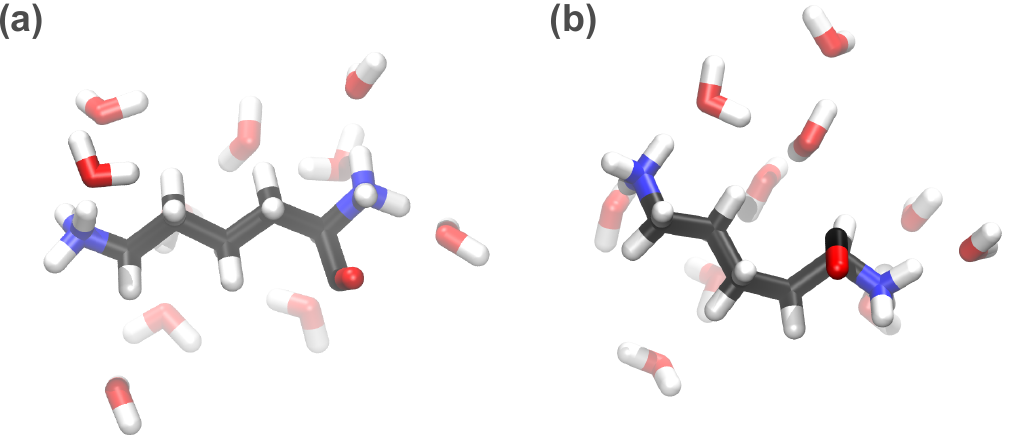}
    \caption{Medoid configurations from clusters (a) 1 and (b) 2 obtained with the Root Mean Square Deviation (RMSD) threshold set to maximize the SS.}
    \label{fig:lysine}
\end{figure}

To investigate the effect of clustering into solvent-related properties, we considered a subset of the lysozyme protein comprising the lysine amino acid and the \num{10} closest water molecules to it.
Since lysine has two oxygen and two nitrogen atoms, as shown in Figure~\ref{fig:lysine}, we computed the number of hydrogen bonds between the amino acid and water molecules for the configurations in each cluster.

\begin{figure}[htbp]
    \centering
    \includegraphics[width=0.95\linewidth]{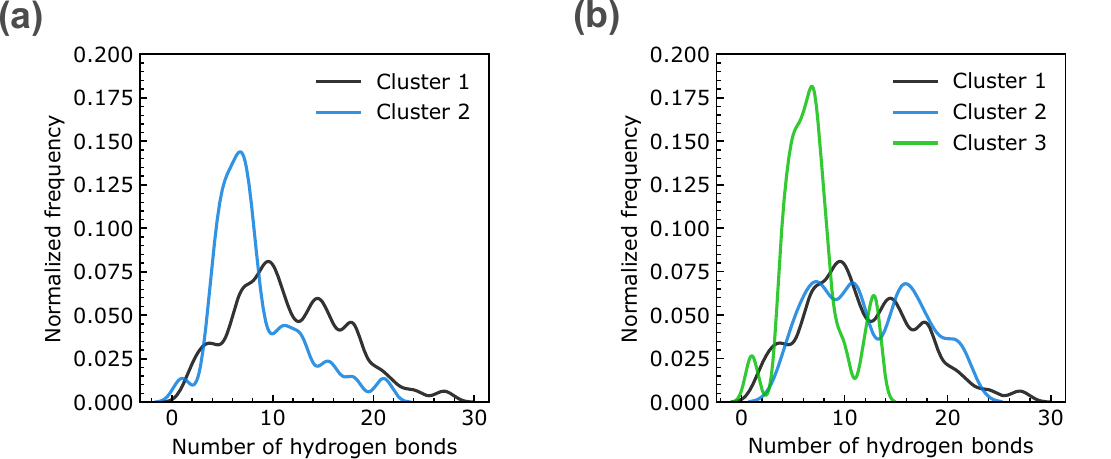}
    \caption{KDE plot of the hydrogen bonds for each cluster with a threshold set to (a) optimize the SS and (b) to \qty{5.5}{\angstrom}.}
    \label{fig:hbonds}
\end{figure}

The Hungarian and the Ward variance minimization methods were employed for the reordering and linkage scheme, respectively, and the total solute weight ($W^\text{solute}$) was set to $0.5$. 
The RMSD threshold was chosen to maximize the silhouette score (SS) \cite{rousseeuw1987} and slightly decreased to \qty{5.5}{\angstrom} to increase the number of clusters from \num{2} to \num{3}. 
Figure~\ref{fig:hbonds} presents the Kernel Density Estimation (KDE) plots \cite{kdeplot1, kdeplot2} of the number of hydrogen bonds for each cluster obtained using both thresholds.

To some extent, the clustering procedure captured differences in the number of hydrogen bonds. 
As shown in Figure~\ref{fig:hbonds} (a), most configurations in Cluster \num{2} have less than \num{10} hydrogen bonds, while Cluster \num{1} has a more even distribution, achieving up to \num{30} H bonds.
One extra cluster is formed when decreasing the threshold through the branching of Cluster \num{2}, as shown in Figure~\ref{fig:hbonds} (b), and it changes the populations from \num{66} and \num{34} for clusters \num{1} and \num{2} to \num{66}, \num{9} and \num{25} for clusters \num{1}, \num{2} and \num{3}, respectively.
Despite the number of hydrogen bonds in configurations from Cluster \num{3} becoming even more localized, the newly formed Cluster \num{2} is similar to Cluster \num{1}, reducing the heterogeneity between clusters.

Further refinement via solute weight tuning and considering different linkage schemes may improve the results.
However, since establishing a hydrogen bond satisfies not only a distance but also an angular criterion \cite{iupac_hbond}, the clustering via RMSD is likely to separate according to the distance but struggles to capture the subtle changes in the angle, especially for larger systems.

\newpage
\section{MOx solvated in methanol}

\begin{figure}[htbp]
    \centering
    \includegraphics[width=0.95\linewidth]{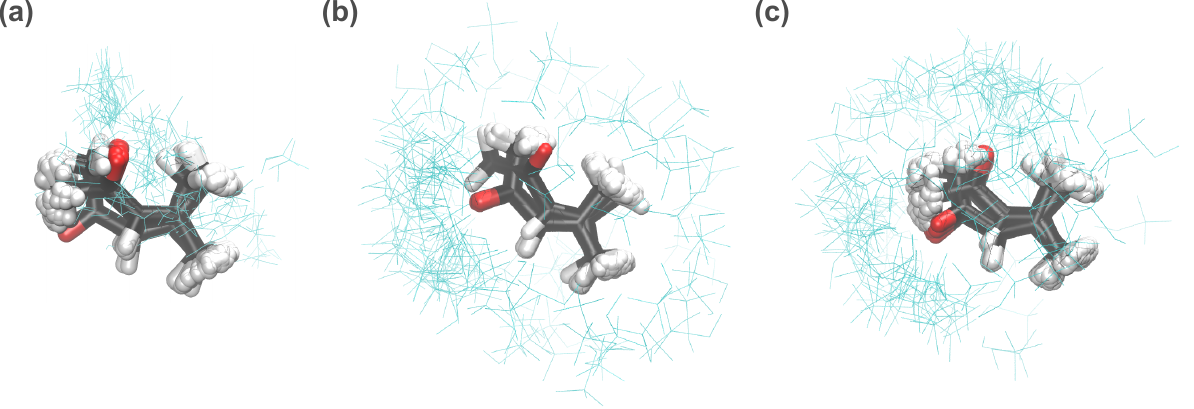}
    \caption{Superposition of MOx configurations for clusters (a) 1, (b) 2 and (c) 3. Methanol molecules are shown as cyan lines for better visualization.}
    \label{fig:mox_SI}
\end{figure}

As mentioned in Section 3.3.1, we performed a clustering procedure for mesityl oxide (MOx) and the four closest methanol molecules with an RMSD threshold of \qty{10}{\angstrom}. 
Considering the same weight for solute and solvent atoms, the Hungarian algorithm was combined with the Ward variance minimization method for the reordering and linkage schemes, respectively. 
As a result, we obtained three clusters and the superposition of configurations from each cluster is shown in Figure~\ref{fig:mox_SI}.

\begin{figure}[H]
    \centering
    \includegraphics[width=0.525\linewidth]{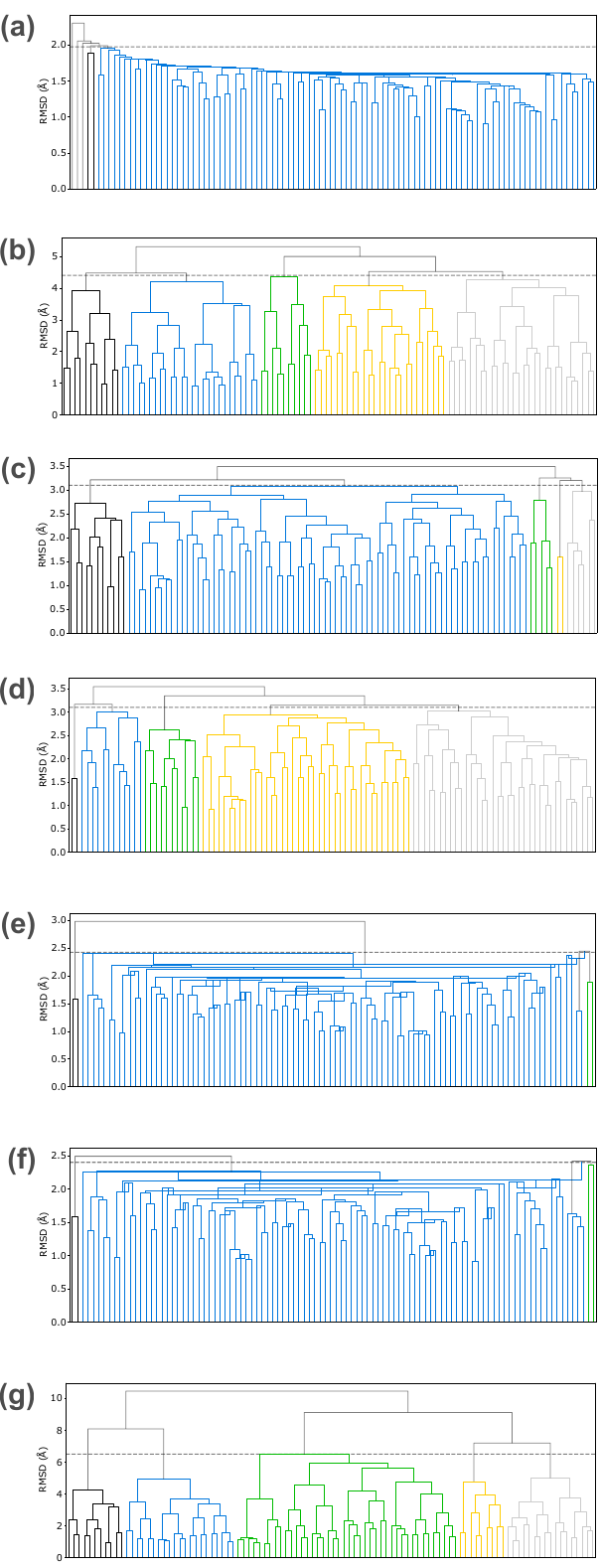}
    \caption{Dendrograms from the clustering of MOx and 4 methanol molecules using (a) single, (b) complete, (c) average, (d) weighted, (e) centroid, (f) median and (g) ward linkage methods.}
    \label{fig:mox_dendro}
\end{figure}

In Section 3.3.1, we also compare different linkage schemes and the corresponding dendrograms are presented in Figure~\ref{fig:mox_dendro}. The visual analysis shows that single, centroid, and median methods fail to form different clusters while increasing the similarity of configurations from the same cluster. This trend is in agreement with the goal of maximizing the CH score. As shown in Table 2, single, centroid and median have a CH score of \num{1.191}, \num{3.186} and \num{3.043}, respectively. On the other hand, average and Ward, the best-performing methods, yielded a Calinski Harabasz (CH) score \cite{harabasz1974} of \num{6.285} and \num{10.110}.

\newpage
\section{Reliability of reordering algorithms}

To compare the quality of the different reordering algorithms, we selected two representative conformations of MOx with four acetonitrile molecules. 
We used the brute force algorithm to reorder the atoms of the solvent molecules and found the optimal solution that minimizes the RMSD to \qty{2.291}{\angstrom}, using the implementation in the \texttt{rmsd} package \cite{Charnley2025}. 
For both configurations, we randomly shuffled the indices of solvent atoms and further scrambled all ``new'' configurations into the same trajectory file. 
Then, we gradually increased the number of identical but shuffled configurations in the trajectory file, and each relabeling algorithm was employed. 
The number of clusters was set to maximize the silhouette coefficient and a final Kabsch rotation was performed. The results are summarized in Table \ref{tab:reordering}. \\

{\renewcommand{\arraystretch}{0.9}\begin{table}[hbp]
    \centering
    \caption{Clustering of randomly shuffled configurations generated from two unique configurations. The performance of each method is compared to trajectory files with 20, 100 and 500 configurations.}
    \label{tab:reordering}
    \begin{tabular}{ccccl}
        \toprule
        Algorithm & \makecell{Number of \\ configurations} & \makecell{Number of \\ clusters} & \makecell{Cluster index :\\size} & RMSD (\AA) : count \\
        \midrule
        \multirow{6}{*}{Hungarian} & \multirow{2}{*}{20} & \multirow{2}{*}{2} & 1 : 10 & 0.000 : 100 \\
        & & & 2 : 10 & 2.314 : 100 \\
        & \multirow{2}{*}{100} & \multirow{2}{*}{2} & 1 : 50 & 0.000 : 2500 \\
        & & & 2 : 50 & 2.314 : 2500 \\
        & \multirow{2}{*}{500} & \multirow{2}{*}{2} & 1 : 250 & 0.000 : 62500 \\
        & & & 2 : 250 & 2.314 : 62500 \\
        \midrule
        \multirow{6}{*}{Distance} & \multirow{2}{*}{20} & \multirow{2}{*}{2} & 1 : 10 & 0.000 : 100 \\
        & & & 2 : 10 & 4.335 : 100 \\
        & \multirow{2}{*}{100} & \multirow{2}{*}{2} & 1 : 50 & 0.000 : 2500 \\
        & & & 2 : 50 & 4.335 : 2500 \\
        & \multirow{2}{*}{500} & \multirow{2}{*}{2} & 1 : 250 & 0.000 : 62500 \\
        & & & 2 : 250 & 4.355 : 62500 \\
        \midrule
        \multirow{6}{*}{QML} & \multirow{2}{*}{20} & \multirow{2}{*}{2} & 1 : 10 & 0.000 : 100 \\
        & & & 2 : 10 & 4.579 : 100 \\
        & \multirow{2}{*}{100} & \multirow{2}{*}{2} & 1 : 50 & 0.000 : 5000 \\
        & & & 2 : 50 & 4.579 : 2500 \\
        & \multirow{2}{*}{500} & \multirow{2}{*}{2} & 1 : 250 & 0.000 : 62500 \\
        & & & 2 : 250 & 4.579 : 62500 \\
        \bottomrule
    \end{tabular}
\end{table}}

Despite being heuristic, the algorithms do not show any dependence on the trajectory size, always converging to the same result. 
The correct number of clusters and the corresponding populations were consistently identified, as the minimized RMSD always converged to the same value, regardless of the atomic indices. 
In the last column of Table \ref{tab:reordering}, we present the unique pairwise RMSD values between two snapshots and the number of times they appear in the RMSD matrix (count).
Given the symmetry of the distance matrix, we only consider the values of the upper triangular part of the RMSD matrix.

As desired, only two values were reported, corresponding to the null RMSD between identical configurations and the non-zero RMSD between different configurations. 
Since the number of shuffled copies of each unique configuration is the same, one should expect equal proportions, as observed.

Comparing the non-zero RMSDs with the reference value allows us to assess the quality of each procedure. 
The Hungarian algorithm achieved the best results with a slight error of \qty{0.023}{\angstrom}, which should have a minor impact on the clustering procedure. 
On the other hand, both the distance and QML algorithms overestimated the RMSD by \qty{2.044}{\angstrom} and \qty{2.288}{\angstrom}, respectively. 
This trend between algorithms was also observed for the examples presented in the main text (see Section 3.4), but when considering a larger set of unique configurations, we found that QML tends to outperform the distance algorithm.
For example, when considering the trajectories with 1 + 10 and 1 + 20 water molecules presented in Section 3.4, the sum of the RMSD matrix elements is \qty{21563}{\angstrom} and \qty{50875}{\angstrom} when using the distance algorithm but \qty{21157}{\angstrom} and \qty{27962}{\angstrom} with the QML algorithm, respectively.
Nevertheless, the quality of heuristic algorithms can be system-dependent and should be carefully investigated for different applications.

\newpage
\section{Illustrative comparison with standard clustering approach}

Considering the same trajectories used in Section S3, we performed an hierarchical clustering of the configurations using the well-established TTClust \cite{ttclust} package. 
Since the trajectories are formed by shuffling the atomic indices of two unique configurations, we set the number of clusters as 2 when running TTClust. 
The results are shown in Table \ref{tab:benchmark} and compared with \texttt{clusttraj} using the Hungarian algorithm. \\

{\renewcommand{\arraystretch}{0.9}\begin{table}[hb]
    \centering
    \caption{Clustering of randomly shuffled configurations generated from two unique configurations with \texttt{clusttraj} and TTClust packages. The performance of each method is compared to trajectory files with 20, 100 and 500 configurations.}
    \label{tab:benchmark}
    \begin{tabular}{ccccc}
        \toprule
        Program & \makecell{Number of \\ configurations} & \makecell{Number of \\ clusters} & \makecell{Cluster index :\\size} & Total time (s) \\
        \midrule
        \multirow{6}{*}{\texttt{clusttraj}} & \multirow{2}{*}{20} & \multirow{2}{*}{2} & 1 : 10 & \multirow{2}{*}{5.23} \\
        & & & 2 : 10 & \\
        & \multirow{2}{*}{100} & \multirow{2}{*}{2} & 1 : 50 & \multirow{2}{*}{8.37} \\
        & & & 2 : 50 & \\
        & \multirow{2}{*}{500} & \multirow{2}{*}{2} & 1 : 250 & \multirow{2}{*}{36.46} \\
        & & & 2 : 250 & \\
        \midrule
        \multirow{6}{*}{TTClust} & \multirow{2}{*}{20} & \multirow{2}{*}{2} & 1 : 12 & \multirow{2}{*}{3.22} \\
        & & & 2 : 8 & \\
        & \multirow{2}{*}{100} & \multirow{2}{*}{2} & 1 : 68 & \multirow{2}{*}{3.28} \\
        & & & 2 : 32 & \\
        & \multirow{2}{*}{500} & \multirow{2}{*}{2} & 1 : 361 & \multirow{2}{*}{4.06} \\
        & & & 2 : 139 & \\
        \bottomrule
    \end{tabular}
\end{table}}

As anticipated, even though we predefined the correct number of clusters, the lack of a reordering scheme resulted in non-physical clusters with varying populations, depending on the trajectory size. 
Since the reordering scheme is the most computationally demanding step, scaling with $\mathcal{O}(N^3)$ for the Hungarian algorithm, where $N$ is the number of atoms from identical molecules (see Section~3.4 of the main text), the overall run time with TTClust is shorter. 
However, despite the computational cost of \texttt{clusttraj}'s approach being higher, the wall time is only a few seconds even for the larger trajectory with \num{500} configurations. 
Considering that these calculations were performed on a commercial laptop, one should be able to benefit from the parallel implementation of \texttt{clusttraj} and cluster configurations from larger trajectories (with more atoms and/or more snapshots) using dedicated high-performance computing resources.

\newpage

\bibliographystyle{plain}
\bibliography{ref}